\documentclass[letterpaper, 10pt, twocolumn]{article}

\usepackage{graphicx}
\usepackage[hmargin=0.7in,vmargin=1in]{geometry}
\usepackage{authblk}

\usepackage[backend=biber,sorting=none]{biblatex}
\usepackage{float}

\usepackage{amsmath,amssymb,amsfonts}%
\usepackage{amsthm}%
\usepackage{gensymb}
\usepackage[symbol]{footmisc}

\usepackage{soul}
\usepackage[dvipsnames]{xcolor}

\newcommand{\fsh}{$f_{\text{SH}}~$}
\newcommand{\fshns}{$f_{\text{SH}}$}

\begin{document}

\author[1,2,*]{Lindell M. Williams}
\author[1]{Grant M. Brodnik}
\author[1,2]{Scott B. Papp}

\affil[1]{Time and Frequency Division, National Institute of Standards and Technology, Boulder, CO 80305, USA}
\affil[2]{Department of Physics, University of Colorado, Boulder, CO 80309, USA}

\title{Second-harmonic stabilization of a bulk photonic resonator}

\date{}
\maketitle
\thispagestyle{empty}

\footnotetext[1]{lindell.williams@colorado.edu}

 \noindent \textbf{ The resonant modes of optical cavities provide a powerful resource for laser-frequency stabilization, underpinning high-precision metrology and coherent signal generation.
Photonic resonators in which the optical mode propagates through material offer a compact alternative to vacuum Fabry-Perot cavity systems, but their performance is limited by sensitivity of the material to the ambient environment.
In this work, we explore second-harmonic (SH) stabilization, which exploits the interplay of a dispersive mode structure against the strict energy conservation of second-harmonic generation. 
Operationally, we use two, 1550 nm lasers to PDH-detect octave-spaced resonant modes of an ultra-high-Q photonic resonator with one laser frequency-doubled to 775 nm.
Under SH stabilization, the microwave frequency offset between the 1550 nm lasers, which we refer to as the SH signal (\fshns) maps the absolute frequency of the 1550 nm laser to an electronic signal.
We characterize this mapping through comparison of the absolute optical frequency inference provided by \fsh to an out-of-loop optical measurement, and our results suggest \fsh accurately proxies frequency drift.
We evaluate the sensitivity and noise floor of this technique, considering contributions from laser locking and bulk material properties, and conclude that \fsh is sufficiently sensitive to enhance long-term laser-frequency stability with respect to the resonator.
These results demonstrate SH stabilization as a useful technique that infers absolute drift, thereby enabling the increased stability of future compact, precision frequency references. }

\maketitle

\section{Introduction}

Optical cavities play a fundamental role in precision metrology, stabilizing laser frequency for applications such as optical atomic clocks \cite{bloom_optical_2014}, pristine microwave signal generation \cite{fortier_generation_2011}, and communications \cite{al-taiy_ultra-narrow_2014}.
Lasers locked to high-finesse, Fabry-Perot cavities feature ultranarrow linewidths and reduced frequency drift \cite{kessler_sub-40-mhz-linewidth_2012, matei_15text_2017, robinson_crystalline_2019, wiens_simplified_2020}, representing the state-of-the-art in short-term stability.
The complicated implementation of these cavities, involving optics, vacuum, and active and passive isolation \cite{boyd_basic_2024}, have motivated new designs and integration.
Photonic resonators in which light propagates through material offer advantages in compactness for applications including gas sensing \cite{liu_cavity_2023} and geodesy \cite{giorgi_advanced_2019}.
These devices span a variety of form factors, including whispering-gallery mode resonators \cite{shitikov_billion_2018, yang_advances_2015, zhang_microrod_2019}, photonic waveguide rings \cite{puckett_422_2021}, and photonic Fabry-Perot resonators \cite{zhang_ultranarrow_2020, zhang_monolithic_2024, cheng_harnessing_2025}, which enable ultrahigh quality factors for narrow $<20$ Hz laser linewidth.
However, such resonators are limited in terms of material availability and volume for enclosure engineering. As a result, compact resonators encounter thermal noise and drift from material sensitivity to ambient exposure, ultimately reaching a thermorefractive-noise-limited linewidth and long-term temperature-induced drift. Indeed, realizing the potential of compact resonators that maintain relatively narrow linewidth requires novel methods of reducing frequency drift.

Record-low frequency drift in optical cavity systems is often achieved by decoupling optical fields from environmental perturbations to reduce resonator sensitivity.
Engineered enclosures isolate Fabry-Perot cavities, utilizing vacuum and thermal controls\textemdash often at cryogenic temperatures \cite{adhikari_cryogenic_2020}\textemdash to create a low-noise environment \cite{kessler_sub-40-mhz-linewidth_2012, matei_15text_2017, robinson_crystalline_2019, wiens_simplified_2020}.
Vacuum \cite{cheng_harnessing_2025}, mono-crystalline materials \cite{kessler_sub-40-mhz-linewidth_2012, robinson_crystalline_2019}, and Fabry-Perot materials and photonic resonators operated at expansion nulls \cite{alnis_subhertz_2008, ito_stable_2017, zhang_cryogenic_2024} all offer compelling options for increased stability. In compact resonators, where size and material constraints limit the use of these techniques, active stabilization based on optical thermometry has been explored. 
Thermometry utilizing orthogonal polarization mode families has been demonstrated in whispering-gallery mode resonators \cite{strekalov_temperature_2011, fescenko_dual-mode_2012, baumgartel_frequency_2012, lim_probing_2019} and photonic waveguide resonators \cite{zhao_integrated_2021}.
In whispering-gallery mode resonators, thermometry using a laser and its second harmonic has also been explored \cite{weng_nano-kelvin_2014}.
However, the complex geometry of whispering-gallery modes can lead to poor correlation between stabilized mode temperature and absolute frequency stability.
With relatively simple geometry and design and demonstrated performance in short-term stability \cite{zhang_ultranarrow_2020}, Fabry-Perot photonic resonators offer an as-of-yet underexplored platform for frequency-drift measurement, using optical thermometry.

We introduce second-harmonic (SH) stabilization, which measures frequency drift in a resonator by balancing dispersion across octave-spaced modes against energy conservation in second-harmonic generation (SHG). For a concrete model of SH stabilization, we consider the mode structure of a dispersive resonator with mode indices, $m$, such that $\nu(m) = \frac{mc}{2n(\nu)L}$, where $L$ is the resonator length and $n(\nu)$ is the refractive index at frequency $\nu$. Performing SH stabilization on a resonator yields a microwave frequency SH signal,

\begin{equation}\label{eq:absolute}
f_{\text{SH}} = \nu_1 - \frac{1}{2}\nu_2 = \nu_1\left(1 - \frac{1}{2} \frac{m_2}{m_1} \frac{n(\nu_1)}{n(\nu_2)}\right),
\end{equation}

\noindent where $\nu_1$ is a resonant frequency at the fundamental and $\nu_2$ is the resonant frequency nearest to its second harmonic ($\nu_2 \approx 2\nu_1$). We note that a small change in \fshns, $\Delta f_\text{SH}$ relates to a corresponding change in absolute resonant frequency, $\Delta \nu_1$, by the total variation of Eq.~\ref{eq:absolute}, $\Delta f_\text{SH} = ( \frac{df_\text{SH}}{d\nu_1})\Delta\nu_1$. Importantly, \fsh is an intrinsic property of the resonator’s dispersive mode structure; once the fundamental mode, $m_1$, is selected, the corresponding second-harmonic mode, $m_2$, is uniquely defined. Therefore, \fsh depends predominantly on resonator material properties, including the frequency-dependent refractive index and thermo-optic coefficient.

Here, we explore SH stabilization in an ultra-high-Q photonic resonator.
We probe \fsh by use of the heterodyne of two lasers; one laser is PDH-locked to the resonator at the fundamental wavelength (1550 nm), and the second is frequency-doubled and PDH-locked at the second harmonic (775 nm).
By the temperature dependence of \fshns, we measure $df_\text{SH}/d\nu_1$, which is consistent with the reference thermo-optic and dispersive properties of the resonator material. Second-harmonic stabilization relies on precise detection of resonator modes, so we analyze the limits of the technique set by both fundamental and excess technical noise in the laser locks. We explore the limits for inferring drift with \fsh and find it can accurately infer drift to the Hz/s level, supporting a three order of magnitude improvement over the free-running photonic resonator. These results establish SH stabilization as a viable method to enhance long-term frequency stability in compact dispersive resonators.

\section{Experimental description}

\begin{figure*}[h]
    \centering
    \includegraphics[width=0.8\linewidth]{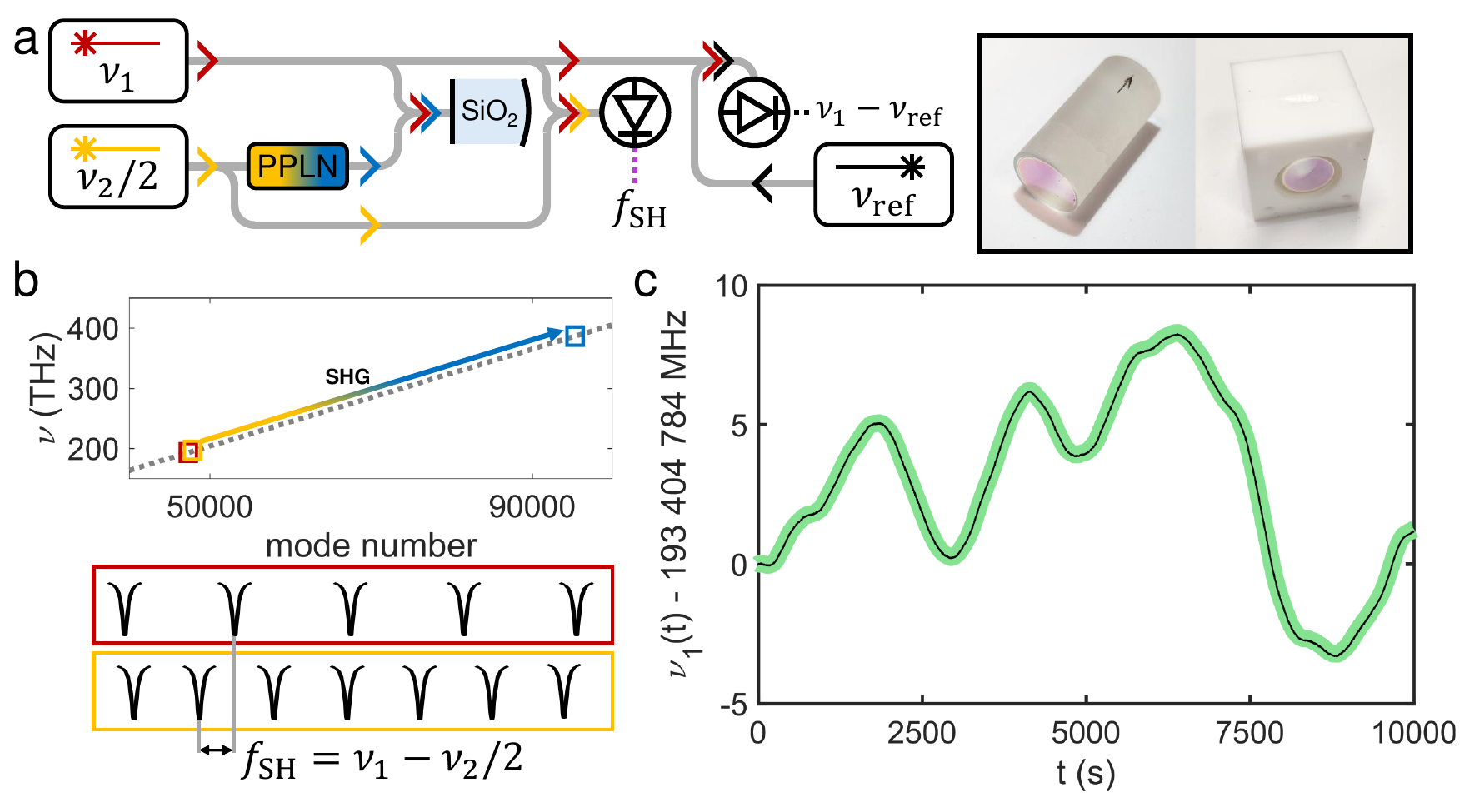}
    \caption{(a) Experimental setup for SH stabilization. One laser at frequency $\nu_1$ directly locks to the dispersive resonator, and the other at frequency $\nu_2/2$ is first frequency-doubled using periodically poled lithium niobate (PPLN) to $\nu_2$ before locking to the resonator.  An additional reference laser, $\nu_\text{ref}$, is used for out-of-loop characterization.  The fused silica Fabry-Perot resonator is shown to the right. (b) Resonant mode structure diagram defining \fsh with modes highlighted around $\nu_1$ (red), $\nu_2/2$ (yellow), and $\nu_2$ (blue). (c) Demonstration of the correlation between changes in \fsh and changes in absolute resonant frequency.  The green band shows resonant frequency fluctuations characterized with the external reference laser, and the black line shows fluctuations in \fsh scaled by a constant factor of 1/($-48.7$).}
    \label{fig:fig1}
\end{figure*}

Figure~\ref{fig:fig1} presents our experiment with SH stabilization, using a Fabry-Perot photonic resonator. The device is a 2.54 cm long, monolithic, fused silica resonator \cite{zhang_ultranarrow_2020, zhang_monolithic_2024} (Fig.~\ref{fig:fig1}a, right) with a plano-convex geometry and a Q of approximately 1 billion.
The resonator is housed in a teflon enclosure designed to provide passive thermal and vibration isolation \cite{zhang_ultranarrow_2020}. 
The resonator and its housing are further enclosed in a small box equipped with PID-controlled resistive heaters set  $15\degree\text{C}$ above ambient temperature to impose a coarse level of thermal stability. We experimentally realize SH stabilization, locking two lasers to the resonator at a given frequency and its second harmonic to measure \fsh (Fig.~\ref{fig:fig1}a, left).
The first laser is directly locked to a resonant mode with frequency $\nu_1$, corresponding to 1550 nm, using the Pound-Drever-Hall (PDH) scheme \cite{drever_laser_1983}.
A second laser at nearly the same frequency is frequency doubled by use of a waveguide-coupled SHG crystal (periodically poled lithium niobate) and locked to an optical mode with resonant frequency $\nu_2$ (775 nm).
By mixing the two lasers on a photodiode, the heterodyne beat, $\nu_1 - \nu_2/2$, realizes \fsh as in Eq.~\ref{eq:absolute}.
We illustrate this definition in Fig.~\ref{fig:fig1}b, showing that the dispersive mode spectrum defines \fsh for a specific fundamental mode and its corresponding second-harmonic mode. We use a commercial stabilized laser with drift of $\sim0.1$ Hz/s as a reference ($\nu_\text{ref}$, 1550 nm) for out-of-loop characterization of our photonic resonator.

Figure~\ref{fig:fig1}c shows our calibration of the scaling factor between fluctuations in \fsh and fluctuations in $\nu_1$.
We simultaneously record \fsh and the out-of-loop reference heterodyne for up to 10,000 s, quantifying $\Delta f_\text{SH}$ and $\Delta \nu_1$ by subtracting the initial frequency from subsequent data.
In Fig.~\ref{fig:fig1}c, we calibrate by fitting the scaled \fsh fluctuations (black line) to resonant frequency fluctuations quantified with the reference heterodyne (green line), finding that the residual is minimized with a scaling of $ \Delta f_\text{SH}= \frac{1}{-48.7}\Delta\nu_1$.
In this work, we treat the scaling factor as constant over time and apply this calibrated scaling to all subsequent inferences of $\nu_1$ from \fshns.

Imprecision and inaccuracy of the resonator material properties, including index of refraction and thermo-optic coefficient, and environmental couplings reduce the absolute accuracy of $\nu_1$ inferred from \fshns.
Nonetheless, we leverage the relationship presented in Eq.~\ref{eq:absolute} by measuring relative changes in \fsh over time.
Operationally, a simultaneous measurement of \fsh and $\nu_1$ yields the partial derivative $\frac{df_{\text{SH}}}{d\nu_1} \approx \frac{\partial f_\text{SH}}{\partial x}\frac{\partial x}{\partial \nu_1}$, where $x$ represents a material property or environmental coupling.
Temperature is a particularly relevant coupling for resonators exposed to the ambient environment, as temperature causes a frequency shift proportional to the thermo-optic coefficient.
A temperature-induced frequency shift of $\nu_1$ results in

\begin{equation}\label{eq:sensitivity}
     \frac{\partial f_\text{SH}}{\partial T}\frac{\partial T}{\partial \nu_1}=  1 - \frac{1}{2} \frac{m_{2}}{m_{1}} \frac{n_1}{n_2}\left( \frac{\frac{1}{n_2}\frac{dn_2}{dT}  + \frac{1}{L} \frac{dL}{dT}}{\frac{1}{n_1}\frac{dn_1}{dT}  + \frac{1}{L} \frac{dL}{dT}} \right),    
\end{equation}

\noindent including frequency-dependence of the thermo-optic coefficient.
The Sellmeier index of refraction for fused silica \cite{malitson_interspecimen_1965} at $\nu_1 = 193~404~610 \pm 20~\text{MHz}$ and $\nu_2 = 386~806~800 \pm 40~\text{MHz}$ and previously reported values for $dn(\nu)/dT$ \cite{leviton_temperature-dependent_2006, rego_temperature_2023} and $(1/L) dL/dT$ \cite{berthold_ultraprecise_1976} predict a scaling of 1/($-48\pm2$), which agrees with our calibration.
Expanding on this result, we note that other environmental conditions can couple to the magnitude of the scaling factor in a similar manner.
Vibration, pressure, and humidity all plausibly have frequency-dependent material couplings analogous to the thermo-optic coefficient \cite{bertholds_determination_1988, sun_optical_2015}, and a similar derivation to Eq.~\ref{eq:sensitivity} can yield those  contributions.
 
\section{Characterization of \fsh}\label{sec:coherence}

\begin{figure*}[h!]
    \centering
    \includegraphics[width=0.8\linewidth]{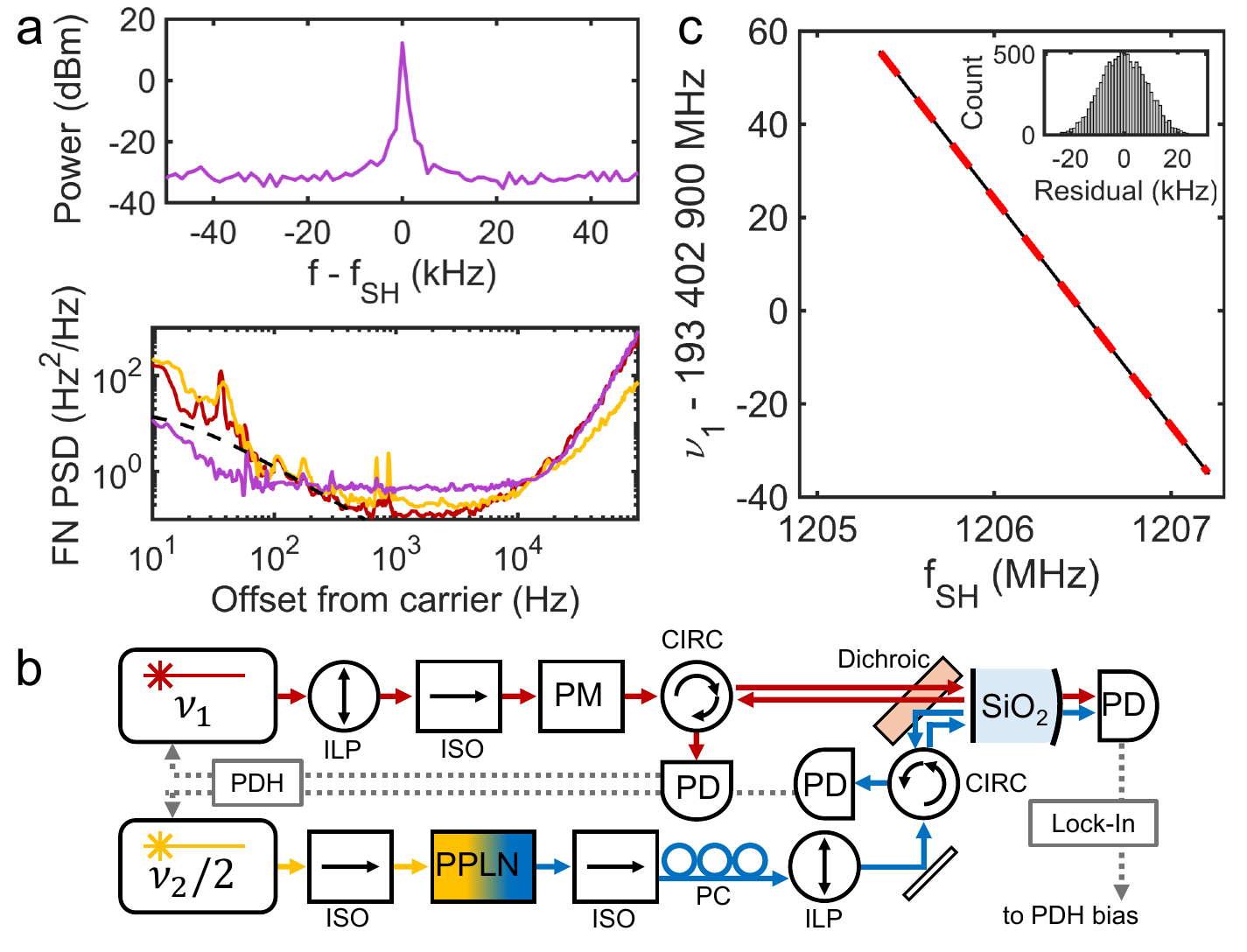}
    \caption{(a) Top: spectrum of \fshns; bottom: frequency noise power spectral density (FN PSD) for $\nu_1$ (red), $\nu_2/2$ (yellow), and \fsh (magenta).  Theoretical estimate of thermal noise is shown in dashed black. (b) System diagram for our experiment.  Gray lines represent feedback loops including PDH locking and transmission lock-in detection.  ILP: in-line polarizer, ISO: isolator, PM: phase modulator, CIRC: circulator, PD: photodetector, PPLN: periodically poled lithium niobate, PC: polarization control. (c) Thermal transient measurement of the scaling factor between \fsh and $\nu_1$, $df_\text{SH}/d\nu_1$.  Linear fit with slope $1/(-48.7 \pm 0.1)$ is shown in dashed red, and residuals to the fit are histogrammed in the upper right inset.}
    \label{fig:fig2}
\end{figure*}

Since \fsh is a microwave signal that directly relates the frequency of our resonator-stabilized laser, we explore the effects of noise and offsets in the PDH measurement systems; see Fig.~\ref{fig:fig2}.
The microwave spectrum of \fsh (Fig.~\ref{fig:fig2}a, top) demonstrates a narrow-linewidth signal with high signal-to-noise ratio for precision frequency metrology.
We examine the frequency noise power spectral density (FN PSD; Fig.~\ref{fig:fig2}a, bottom) of \fsh (magenta) and its two lasers (red, yellow).
Using the PSD, we calculate $1/\pi$ integral linewidths for \fshns, $\nu_1$, and $\nu_2/2$ to be 16 Hz, 35 Hz, and 36 Hz, respectively. 
At $10^2-10^3~\text{Hz}$ offset from carrier, the individual lasers feature frequency noise with a PSD of $10^0-10^1~\text{Hz}^2/\text{Hz}$ due to thermorefractive noise of the photonic resonator, consistent with photonic resonator thermal noise \cite{zhang_microrod_2019, zhang_ultranarrow_2020}.
For offset frequency above 400 Hz, the white frequency noise spectrum of the PDH-locked lasers is approximately $4\times 10^{-1}~\text{Hz}^2/\text{Hz}$.
On \fshns, we see the quadrature sum of the lasers' white noise.
The white noise therefore sets a fundamental noise floor in our implementation of this technique, although residual technical noise limits our results in practice.
Using \fsh to infer frequency fluctuations in $\nu_1$ requires multiplying by the reciprocal of the scaling factor, $1/(df_\text{SH}/d\nu_1)$; when \fsh is scaled by our calibrated factor, the white noise floor becomes $\sim 10^3~\text{Hz}^2/\text{Hz}$.
Laser dynamics with a frequency noise PSD below this level are impossible to infer from \fshns.
Because high-offset-frequency noise of the laser falls below this level, our implementation of this technique cannot infer the short-term stability of the resonator-stabilized laser.

A long-duration measurement of \fsh is susceptible to residual technical noise that degrades the accuracy of drift inference.
We further detail our experimental setup with SH stabilization, highlighting the methods used to minimize residual technical noise that affects \fshns, in Fig.~\ref{fig:fig2}b.
We use polarization-maintaining (PM) fibers and PM-fiber-coupled phase modulators, splitters, isolators, in-line polarizers and circulators to maintain linear polarization with respect to the phase modulator crystal axis and the birefringent polarization states of the resonator \cite{zhang_ultranarrow_2020}.
We measure the PDH bias before and after our measurements and observe that the bias shifts slowly over long measurements.
We attribute this bias shift to residual amplitude modulation (RAM) \cite{wong_servo_1985, whittaker_residual_1985, zhang_reduction_2014}, which is introduced when polarization noise from the phase modulators used for PDH is passed through polarizing components.
Since RAM serves to change the PDH lock bias, we expect transmission through the cavity to vary as the lasers' frequencies change relative to true resonance minima.
We include additional lock-in detection on the transmission through the resonator as a slow feedback loop to combat this effect and keep the PDH-locking loop centered on the true resonance minimum.
With this, we see a reduction in PDH bias drift, however it still remains at a small but significant level.
While our system includes efforts to minimize RAM-induced PDH bias noise, future work on SH stabilization could aim to further eliminate RAM to realize more accurate measurements of \fshns.

Technical noise affects the ability to infer $\nu_1$ using \fsh at a level depending on $df_{\text{SH}}/d\nu_1$.
We directly measure $df_{\text{SH}}/d\nu_1$ by adjusting the resonator enclosure temperature setpoint and simultaneously recording the frequencies of \fsh and the out-of-loop reference heterodyne during equilibration.
By changing the temperature, we establish conditions in which thermal gradients dominate over pressure, humidity, and vibrations, thereby satisfying the assumption of a temperature-induced frequency change made in Eq.~\ref{eq:sensitivity}.
We apply a change of $+0.1 \degree\text{C}$, recording a frequency shift of approximately $-$100 MHz for $\nu_1$ and +2 MHz for \fsh over 1,000 s.
We plot this in Fig.~\ref{fig:fig2}c alongside a linear fit and a histogram of the fit residuals.
Using this fit, we determine $df_{\text{SH}}/d\nu_1$ to be $1/(-48.7 \pm 0.1)$.
The uncertainty in this measurement is dominated by the effects of technical noise in the system, including RAM, but also includes the influences of other environmental effects not accounted for in the approximation $\frac{df_{\text{SH}}}{d\nu_1} \approx \frac{\partial f_\text{SH}}{\partial T}\frac{\partial T}{\partial \nu_1}$.
We find the correlation to be exceptionally linear ($R^2 > 1-10^{-7}$), suggesting that the further reduction in these remaining sources of uncertainty can lead to a significantly more precise measurement of the scaling factor.
This measurement is consistent with the calibrated value found in Fig.~\ref{fig:fig1}c for steady-state exposure, suggesting that our ambient conditions sufficiently satisfy the assumptions of Eq.~\ref{eq:sensitivity}.

\section{Frequency Drift Inference}

\begin{figure*}[h]
    \centering
    \includegraphics[width=0.8\linewidth]{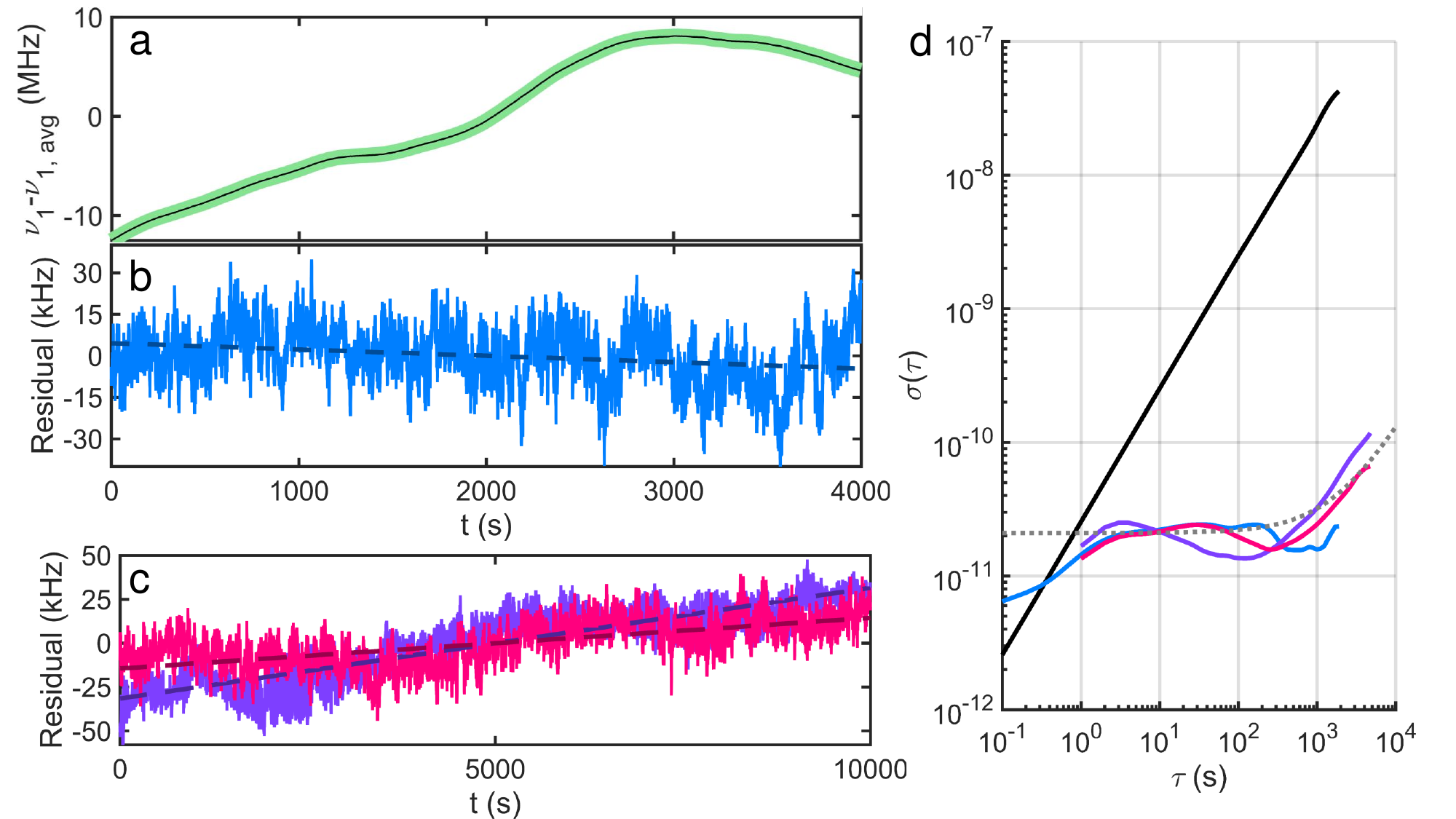}
    \caption{(a) Comparison of true optical deviation derived from external reference laser (green band) and scaled \fsh (black line). (b) Residual of truth signal and scaled \fshns.  Dashed line shows linear fit for residual frequency drift. (c) Residuals for two longer data sets and their resulting linear drift fits.  (d)  Allan deviation for original optical signal shown in black.  Allan deviations for the residual signals in panels b and c are plotted in their respective colors.  An approximate power-law summary is include in dashed gray featuring 3 Hz/s linear drift and flicker noise that contributes $(-48.7)^2 \times 5\times10^3~\text{Hz}^2/\text{Hz}$ to the frequency noise PSD at 1 Hz offset.}
    \label{fig:fig3}
\end{figure*}

To explore the applicability of SH stabilization in improving long-term stability, we measure the accuracy of inferring frequency drift with \fshns; see Fig.~\ref{fig:fig3}.
We simultaneously record \fsh and the out-of-loop reference heterodyne during steady-state environmental conditions for times up to 10,000 s (Fig.~\ref{fig:fig3}a).
To infer fluctuations in $\nu_1$, we scale fluctuations in \fsh by our calibrated factor of 1/($-$48.7).  We then take the difference between this scaled \fsh and the reference heterodyne to create a residual signal that we use to characterize the inference's accuracy.

We find that SH stabilization infers drift to the Hz/s level, with further precision limited by the residual technical noise in our system.
We plot experimental data for a measurement period of 4,000 s in Fig.~\ref{fig:fig3}a.
We plot the corresponding residual signal in Fig.~\ref{fig:fig3}b and include a linear fit as a characterization of any residual drift.
We see that although short-term fluctuations are increased by the scaling factor, the drift of 6.5 kHz/s is accurately inferred with the residual featuring a drift of only $-$2.3 Hz/s.
We also include the residual signal for two longer measurements of 10,000 s in Fig.~\ref{fig:fig3}c.
For these longer measurements, we find a residual frequency drift of 7 Hz/s (purple) and 3 Hz/s (pink).
Based on our system characterization, we attribute this level of residual drift to the effects of RAM.
As the phase modulators and PM fiber slowly change temperature, the magnitude of RAM affecting each laser varies.
Accordingly, the changing PDH bias offsets cause \fsh to exhibit a slow frequency drift unrelated to resonator dynamics.
Without RAM, we expect the long term stability to be limited by the scaled flicker floor arising from technical noise in laser stabilization electronics, which dominates the noise PSD at low frequency offsets.
We estimate this flicker noise to contribute $5\times10^3~\text{Hz}^2/\text{Hz}$ to the frequency noise PSD of \fsh at 1 Hz offset, which corresponds to a fractional frequency stability of the resonator-stabilized laser of $2\times10^{-11}$ at all averaging times after scaling by our calibrated factor.

To summarize fluctuations in $\nu_1$ and the inference provided by \fshns, we plot the Allan deviation of the free-running laser and the residual signals in Fig.~\ref{fig:fig3}c.
Over the plotted range of offset frequencies, the Allan deviation of the free-running resonator (black line) features $\tau^1$ dependence characteristic of frequency drift with a magnitude of 6.5 kHz/s.
At averaging times less than $300$ s, the stability of the residual signals (colors corresponding to Fig.~\ref{fig:fig3}b,c) is limited by the scaled flicker noise, while the residual drift begins to dominate at longer averaging times.
At an averaging time of 1,000 s, the residual signals exhibit fractional frequency stability in the range of 1\textendash 3 $\times 10^{-11}$.
Compared to the free-running resonator, which has a stability of $2\times10^{-8}$ at 1,000 s, this represents a three order of magnitude improvement.
In Fig.~\ref{fig:fig3}, we include a power-law summary of different experimental measurements (gray dashed line), featuring a scaled flicker floor with frequency noise PSD of $(48.7)^2 \times5\times10^3~\text{Hz}^2/\text{Hz}$ at 1 Hz offset and a residual linear drift of 3 Hz/s.
We omit the limit set by scaled white noise in this plot because it falls significantly below the effects of flicker noise and remaining drift, although it would pose an ultimate limit should these other contributions be reduced.

\section{Conclusion}

We have demonstrated SH stabilization as a technique that accurately infers frequency drift in a dispersive resonator to the Hz/s level.
Leveraging energy conservation in SHG, our system realizes SH stabilization using two lasers locked to the resonator.
We show that the fluctuations in \fsh accurately infer corresponding fluctuations in absolute resonant frequencies, related by a material-defined scaling factor.
We both calibrate and directly measure this factor, finding a consistent value that agrees with a theoretical estimate leveraging previously reported properties of the resonator material.
We also examine the limits of SH stabilization in its inference of resonant frequency fluctuations by analyzing the effects of fundamental and excess sources of noise, including thermo-refractive noise, shot noise, and RAM introduced in the PDH-locking scheme.
Because the residual drift and uncertainty in the scaling factor are consistent with the effects of RAM, we identify the further elimination of RAM as a focal point in future precision experiments involving SH stabilization.

Another avenue for further exploration into SH stabilization is in applying the frequency drift inference for active frequency control, e.g. feedforward or feedback.
We first note that, while applying the scaling factor to infer drift from a measured \fsh is in-loop and does not require an independent reference, the calibration of the scaling factor we performed does require a more stable reference.
One possibility for a system that applies SH stabilization would be to undergo an initial calibration period to determine the scaling factor, analogous to the calibration in Fig.~\ref{fig:fig1}c or the direct measurement in Fig.~\ref{fig:fig2}c.
Such an implementation could suffer from inaccuracy if environmental conditions shift and the actual scaling factor changes from the calibrated value, likely prompting recalibration if the changes are significant and persistent.
Perhaps the most compelling method of applying SH stabilization would be in the case of an in-situ, external-reference-free measurement of the scaling factor.
This could involve modulating the cavity and examining how \fsh responds, thus conveying the scaling factor.
Such a method to modulate the cavity in a known fashion without relying on an external reference remains an open challenge, but a solution could herald the robust and practical implementation of SH stabilization.

Looking forward, SH stabilization provides a compelling technique to infer drift in dispersive resonators.
Its application to other form-factors and further application to compact Fabry-Perot geometries could serve to progress practical and scalable resonators to provide stable absolute frequencies with long-term operation in ambient conditions.
Such advancement would support the evolution of precision metrology in an increasingly wide array of environments.

\clearpage
\printbibliography

\clearpage

\section{Methods}

\noindent \textbf{Measurements of \fsh}

\noindent We measure \fsh by first passing the microwave heterodyne signal produced by a photodiode through a frequency divider, dividing the $\sim1$ GHz \fsh by 32 so that it is within the bandwidth of our frequency counter (Keysight 53230a).
We synchronize two counters to measure both the $\nu_1-\nu_\text{ref}$ (also divided) and \fsh heterodynes simultaneously.
For our measurements of 10,000 ($<$10,000) s, we use a gate time of 1 (0.1) s externally triggered by a waveform generator for synchronization between the two counters; the dead-time between gating periods is 10 (100) $\mu$s.
\\

\noindent \textbf{Transmission Lock-in Detection}

\noindent In the free-space path after transmission through the resonator, we include a dichroic beamsplitter to separate transmission of $\nu_1$ and $\nu_2$ and focus them each onto a photodetector.
We apply a lock-in modulation to the DC bias of our PID controllers used for PDH locking.
We choose lock-in modulation frequencies that are significantly faster than the apparent drift caused by RAM, which we observe on the order of 1000 s, while also being significantly slower than the bandwidth of the PID controllers.
We use two different modulation frequencies for $\nu_1$ and $\nu_2$ that are between 100 and 200 Hz, avoiding integer multiples of 60 Hz.
By using a slow feedback loop from the lock-in amplifier (SRS SR830 DSP) to maximize the transmission amplitude, the laser is, on average, centered on the cavity resonance despite possible DC bias offsets caused by RAM.
Although the lock-in modulation means the laser may instantaneously be off-resonance, the relatively high modulation frequency and its single-frequency nature means we can still accurately infer drift using the the average laser frequencies.

\section{Acknowledgements}

We thank Aidan Jones and Sarang Yeola for careful review of the manuscript.
\\

\noindent \textbf{Funding:} This research has been funded by AFOSR FA9550-20-1-0004 Project Number 19RT1019, NSF Quantum Leap Challenge Institute Award OMA – 2016244, and NIST.
\\

\noindent \textbf{Competing interests:} The authors declare no competing interests. This work is a contribution of the US Government and is not subject to US copyright. Mention of specific companies or trade names is for scientific communication only and does not constitute an endorsement by NIST.
\\

\noindent \textbf{Data and materials availability:} All data necessary to support the claims within the paper are included in the plots. Reasonable requests for numeric data will be honored.

\end{document}